\documentclass[12pt]{article}
\usepackage{amsmath} 
\input{epsf} 
\def\ltap{\raisebox{-.6ex}{\rlap{$\,\sim\,$}} \raisebox{.4ex}{$\,<\,$}} 
\def\gtap{\raisebox{-.6ex}{\rlap{$\,\sim\,$}} \raisebox{.4ex}{$\,>\,$}}

\newcommand\as{\alpha_{\mathrm{S}}} 
\newcommand\f[2]{\frac{#1}{#2}}

\def\beq{\begin{equation}} 
\def\eeq{\end{equation}} 
\def\beeq{\begin{eqnarray}} 
\def\eeeq{\end{eqnarray}} 
\def\bom#1{{\mbox{\boldmath $#1$}}} 
\def\to{\rightarrow}
 
\def\nn{\nonumber} 
 
\def\qt{q_T} 

\def\ms{${\overline {\rm MS}}$}

\def\b0{b_0}

\def\tL{{\widetilde L}}

\begin{document} 

\begin{titlepage}
\renewcommand{\thefootnote}{\fnsymbol{footnote}}
\begin{flushright}
SFB/CPP-07-18 \\
KA-TP-13-2007 
     \end{flushright}
\par \vspace*{4mm}

\begin{center}

{\Large \bf
Higgs boson production at the LHC:\\[0.2cm]
transverse-momentum resummation\\[0.4cm]
and rapidity dependence}
 
\end{center}
\par \vspace*{1mm}
\begin{center}
{\bf Giuseppe Bozzi${}^{(a)}$, 
Stefano Catani${}^{(b)}$,}\\ 
\vskip .2cm
{\bf Daniel de Florian${}^{(c)}$
and
Massimiliano Grazzini${}^{(b)}$}\\

\vspace{5mm}

${}^{(a)}$Institut f\"ur Theoretische
Physik, Universit\"at Karlsruhe,\\
P.O. Box 6980, D-76128 Karlsruhe, Germany
\\

\vspace{2mm}

${}^{(b)}$INFN, Sezione di Firenze and Dipartimento di Fisica,
Universit\`a di Firenze,\\ 
I-50019 Sesto Fiorentino, Florence, Italy\\

\vspace{2mm}

${}^{(c)}$Departamento de F\'\i sica, Facultad de Ciencias Exactas 
y Naturales, Universidad de \\ Buenos Aires,
(1428) Pabell\'on 1 Ciudad Universitaria, Capital Federal, Argentina\\

\vspace*{3mm}

\end{center}

\par \vspace{1mm}
\begin{center} {\large \bf Abstract} \end{center}
\begin{quote}
\pretolerance 10000
We consider Higgs boson production by gluon fusion in hadron collisions.
We study the doubly-differential 
transverse-momentum $(q_T)$ and rapidity $(y)$
distribution of the Higgs boson in perturbative QCD.
In the region of small $q_T$
$(q_T \ll M_H$, $M_H$ being the mass
of the Higgs boson), we include the effect of logarithmically-enhanced
contributions due to
multiparton radiation to all perturbative orders.
We use the impact parameter and double Mellin moments to implement and
factorize the multiparton kinematics constraint of transverse- and
longitudinal-momentum conservation. The logarithmic terms are then
systematically resummed in exponential form.
At small $q_T$, we perform 
the all-order resummation of large logarithms 
up to next-to-next-to-leading logarithmic accuracy,
while at large $q_T$ ($q_T \sim M_H$), we apply a matching procedure that
recovers the fixed-order perturbation 
theory up to next-to-leading order.
We present quantitative results for the differential
cross section 
in $q_T$ and $y$
at the LHC,
and we comment on 
the comparison with the $q_T$ cross section integrated over $y$.
\end{quote}

\vspace*{\fill}
\vspace*{2mm}
\begin{center} 
{\large \it This paper is dedicated to the memory of} \\
\vspace*{2mm}
{\large \it Jiro Kodaira, great friend and distinguished colleague }
\end{center}
\vspace*{3mm}
\begin{flushleft}
May 2007

\end{flushleft}
\end{titlepage}

\setcounter{footnote}{1}
\renewcommand{\thefootnote}{\fnsymbol{footnote}}

\section{Introduction}
\label{sec:intro}

The search for the Higgs boson \cite{Hrev}
and the study of its properties (mass, couplings,
decay widths) at hadron colliders require a detailed understanding of
its production mechanisms. This demands reliable computations of related 
quantities, such as production cross sections and the associated distributions 
in rapidity and transverse momentum. In this paper we consider the production 
of the Standard Model (SM) Higgs boson
by the gluon fusion mechanism.

The gluon fusion process $gg \to H$, through a heavy-quark (mainly, top-quark) 
loop, is the main production mechanism of the 
SM Higgs boson $H$ at hadron colliders. 
When combined with the decay channels 
$H \to \gamma \gamma$ and $H \to ZZ$,
this production mechanism is one of 
the most important for Higgs boson searches and studies over the entire
range, 100~GeV$\ltap M_H \ltap$1~TeV, 
of Higgs boson mass $M_H$
to be investigated at the LHC
\cite{atlascms}. 
In the mass range 140~GeV$\ltap M_H \ltap$180~GeV, the gluon fusion process,
followed by the decay $H \to WW \to \ell^+ \ell^- \nu {\bar \nu}$, can be
exploited as main discovery channel
at the LHC and also at the Tevatron \cite{Carena:2000yx}, provided
the background from $t {\bar t}$ production is suppressed by applying a veto cut
on the transverse momenta of the jets accompanying the final-state leptons.

The dynamics of the gluon fusion mechanism in controlled by 
strong interactions. Detailed studies of the effect of QCD radiative 
corrections are thus necessary 
to obtain accurate theoretical predictions.

In QCD perturbation theory, the leading order (LO) contribution
to the total cross section for Higgs boson production by gluon fusion
is proportional to $\as^2$, $\as$ being the QCD coupling.
The QCD radiative corrections to the total cross section are known at the
next-to-leading order (NLO) 
\cite{Dawson:1991zj}-\cite{Muhlleitner:2006wx}
and at the next-to-next-to-leading order (NNLO) 
\cite{Harlander:2000mg}-\cite{NNLOtotal}.
The Higgs boson rapidity distribution is also known at the NLO
\cite{Anastasiou:2002qz} and at the NNLO \cite{Anastasiou:2005qj,Catani:2007vq}. 
The effects of a jet veto have been studied up to the NNLO
\cite{Catani:2001cr,Anastasiou:2005qj,Catani:2007vq}.
We recall that all the results at NNLO have been obtained by using 
the large-$M_t$ approximation, $M_t$ being the mass of the top quark.
This approximation is justified by the fact that the bulk of the QCD 
radiative corrections to the total cross section
is due to virtual and soft-gluon contributions
\cite{Kramer:1996iq,Catani:2001ic,Harlander:2001is,Catani:2001cr,Catani:2003zt}.
The soft-gluon dominance also implies that higher-order perturbative
contributions can reliably be estimated by applying resummation methods 
\cite{Catani:2001ic} of {\em threshold logarithms}, a type of
logarithmically-enhanced terms 
due to multiple soft-gluon emission.
In Ref.~\cite{Catani:2003zt}, the NNLO calculation of the total cross section
is supplemented with threshold resummation at the next-to-next-to-leading logarithmic
(NNLL) level; the residual perturbative uncertainty at the LHC 
is estimated to be at the level of better than $\pm 10$\%. 
The NNLL+NNLO results \cite{Catani:2003zt}
are nicely confirmed by the more recent computation 
\cite{Moch:2005ky}-\cite{Idilbi:2005ni}
of the soft-gluon terms at N$^3$LO; the quantitative effect \cite{Moch:2005ky}
of the additional (i.e., beyond the NNLL order)
single-logarithmic term at N$^3$LO is consistent with the estimated uncertainty
at NNLL+NNLO.
The effect of threshold logarithms on the rapidity distribution of
the Higgs boson has been considered in Ref.~\cite{Ravindran:2006bu}.

The gluon fusion mechanism at ${\cal O}(\as^2)$ 
produces a Higgs boson with a vanishing transverse momentum $\qt$. A large
(or, however, non-vanishing) value of $\qt$ can be obtained only starting from 
${\cal O}(\as^3)$, when the Higgs boson is accompanied by at least one recoiling
parton in the final state. This mismatch by a power of $\as$ is a preliminary
indication of the fact that the small-$q_T$ and large-$q_T$ regions are
controlled by different dynamics regimes.

The large-$q_T$ region is identified by
the condition $q_T \sim M_H$.
In this region, the perturbative series is controlled by a small expansion
parameter, $\as(M_H^2)$, and calculations based on the truncation of the series
at a fixed order in $\as$ are theoretically justified.
The LO, i.e. ${\cal O}(\as^3)$, calculation is reported
in Ref.~\cite{Ellis:1987xu}. The results of 
Ref.~\cite{Ellis:1987xu} and the higher-order studies 
of Refs.~\cite{DelDuca:2001fn,Smith:2005yq}
show that the large-$M_t$
approximation is sufficiently accurate
also in the case of the $q_T$ distribution when $q_T \ltap M_H$,
provided $q_T \ltap M_t$.
Using the large-$M_t$ approximation, the NLO QCD computation
of the $q_T$ distribution of the SM Higgs boson
is presented in 
Refs.~\cite{deFlorian:1999zd,Ravindran:2002dc,Glosser:2002gm,
Anastasiou:2005qj,Catani:2007vq}.
QCD corrections beyond the NLO are evaluated in Ref.~\cite{deFlorian:2005rr},
by implementing threshold 
resummation at the next-to-leading logarithmic (NLL) level.
The results of the numerical programs of 
Refs.~\cite{Anastasiou:2005qj,Catani:2007vq} can also be safely
(i.e. without encountering infrared divergences) extended from large values of
$q_T$ to $q_T=0$: in the small-$q_T$ region these programs evaluate the 
$q_T$ distribution up to NNLO.

In the small-$q_T$ region ($q_T\ll M_H$), where the bulk of events is produced,
the convergence of the fixed-order expansion is definitely spoiled, since
the coefficients of the perturbative series in $\as(M_H^2)$ are enhanced
by powers of large logarithmic terms, $\ln^m (M_H^2/q_T^2)$. The logarithmic
terms are produced by multiple emission of soft and collinear partons (i.e.
partons with low transverse momentum).
To obtain reliable perturbative predictions, these terms have to be 
resummed to all orders in $\as$.
The method to systematically perform all-order resummation of
classes of logarithmically-enhanced terms at small $q_T$ is known
\cite{Dokshitzer:hw}-\cite{Catani:2000vq}.
In the case of the SM Higgs boson, 
resummation has been explicitly worked out at
leading logarithmic (LL), 
NLL \cite{Catani:vd,Kauffman:cx} and 
NNLL \cite{deFlorian:2000pr} level.

The fixed-order and resummed approaches at small and large values of 
$q_T$ can then be matched at intermediate values of $q_T$,
to obtain QCD predictions for the entire range of transverse momenta. 
Phenomenological studies of the SM Higgs boson
$q_T$ distribution at the LHC have been performed in 
Refs.~\cite{Balazs:2000wv}-\cite{Cao:2007du},
by combining resummed and fixed-order perturbation theory at different levels
of theoretical accuracy. 
Other recent studies of various kinematical distributions
of the SM Higgs boson at the LHC are presented in
Refs.~\cite{Gawron:2003np}-\cite{Anastasiou:2007mz}.

In Refs.~\cite{Bozzi:2003jy,Bozzi:2005wk} we studied 
the Higgs boson $q_T$ distribution integrated over the rapidity.
In the small-$q_T$ region, the logarithmic terms were 
systematically resummed in exponential form by working in impact-parameter 
and Mellin-moment space. 
A constraint of perturbative unitarity was imposed on the resummed terms,
to the purpose of reducing the effect of unjustified higher-order 
contributions at large values
of $q_T$ and, especially, at intermediate values of $q_T$.
This constraint thus decreases the uncertainty in the matching procedure 
of the resummed and fixed-order contributions.
Our best theoretical predictions were obtained by matching 
NNLL resummation at small $q_T$ and NLO perturbation theory at large $q_T$.
NNLL resummation includes the complete NNLO result at small $q_T$,
and the unitarity constraint assures that the total cross section at NNLO
is recovered upon integration
over $q_T$ of the transverse-momentum spectrum. 
Considering SM Higgs boson production at the LHC, we concluded 
\cite{Bozzi:2005wk} that the residual perturbative QCD uncertainty of the 
NNLL+NLO result is uniformly of about $\pm 10$\%
from small to intermediate 
values of transverse momenta.

In this paper we extend our study to include the dependence on the
rapidity of the Higgs boson. Using the impact parameter and 
{\em double} Mellin moments, we can perform the extension by maintaining
all the main features of the resummation formalism of 
Refs.~\cite{Catani:2000vq,Bozzi:2005wk}. We are then able to present results
up to NNLL+NLO accuracy
for the doubly-differential cross section 
in $q_T$ and rapidity at the LHC.

The paper is organized as follows. In Sect.~\ref{sec:genfor} we recall
the main aspects of the resummation formalism, and we illustrate the steps that
are necessary to include the dependence on the rapidity in the
$\qt$ resummed formulae.
In Sect.~\ref{sec:lhc} we apply the formalism to
the production of the SM Higgs boson at the LHC, and we perform quantitative 
studies on the $q_T$ and rapidity dependence of the 
doubly-differential cross section. Some concluding remarks are presented in 
Sect.~\ref{sec:fin}. Additional technical details on the 
double Mellin moments of the resummation formulae are given in 
Appendix~\ref{appa}.

\section{Rapidity dependence in $q_T$ resummation}
\label{sec:genfor}

We consider the inclusive hard-scattering process
\begin{equation}
\label{process}
h_1(p_1) + h_2(p_2) \to H(y,\qt,M_H) + X \;\;,
\end{equation}
where the collision of the two hadrons $h_1$ and $h_2$ with
momenta $p_1$ and $p_2$ produces the Higgs boson $H$,
accompanied by an arbitrary and undetected final state $X$.
The centre-of-mass energy of the colliding hadrons is denoted by $\sqrt s$.
The rapidity, $y$, of the Higgs boson is defined in the centre-of-mass frame
of the colliding hadrons, and the forward direction ($y > 0$) is identified by 
the direction of the momentum $p_1$.

According to the QCD factorization theorem,
the doubly-differential cross 
section for this process is
\begin{eqnarray}
\f{d\sigma}{dy \, dq_T^2}(y,q_T,M_H,s) &=& \sum_{a_1,a_2}
\int_0^1 dx_1 \,\int_0^1 dx_2 \; f_{a_1/h_1}(x_1,\mu_F^2)
\,f_{a_2/h_2}(x_2,\mu_F^2) \nn \\
\label{dcross}
&\times&
\f{d{\hat \sigma}_{a_1a_2}}{d{\hat y} \,dq_T^2}({\hat y},q_T, M_H,{\hat s};
\as(\mu_R^2),\mu_R^2,\mu_F^2) 
\;,
\end{eqnarray}
where $f_{a/h}(x,\mu_F^2)$ ($a=q_f,{\bar q_f},g$) are the parton densities of 
the colliding hadrons at the factorization scale $\mu_F$,
$d{\hat \sigma}_{ab}$ are the
partonic cross sections, and $\mu_R$ is the renormalization scale.
Throughout the paper we use parton
densities as defined in the \ms\
factorization scheme, and $\as(q^2)$ is the QCD running coupling in the \ms\
renormalization scheme. The rapidity, $\hat y$, and the centre-of-mass energy,
${\hat s}$, of the partonic cross section (subprocess) are related to
the corresponding hadronic variables $y$ and $s$:
\begin{equation}
\label{kin}
{\hat y} = y - \frac{1}{2} \ln\frac{x_1}{x_2} \;, \quad 
\quad {\hat s}=x_1x_2s \;\;,
\end{equation}
with the kinematical boundary $|{\hat y}| < \ln \sqrt{{\hat s}/M^2}$
$(\, |y| < \ln \sqrt{s/M^2} \,)$ and ${\hat s} > M^2$ $(s > M^2)$.

The partonic cross section $d{\hat \sigma}_{ab}$ is computable in QCD 
perturbation theory. Its power series expansion in $\as$ contains  
the logarithmically-enhanced terms, $(\as^n/q_T^2)\, \ln^m (M_H^2/q_T^2)$,
that we want to resum. To this purpose, we use the general 
(process-independent) strategy and the formalism described in detail in
Ref.~\cite{Bozzi:2005wk}. The only difference with respect to 
Ref.~\cite{Bozzi:2005wk} is that the resummation is now performed at fixed 
values of the rapidity $y$, rather than after integration over the rapidity 
phase space. In the following we briefly recall the main steps of the 
resummation formalism, and we point out explicitly the differences with 
respect 
to Ref.~\cite{Bozzi:2005wk}.

We first rewrite (see Sect.~2.1 in Ref.~\cite{Bozzi:2005wk}) 
the partonic cross section as the sum of two terms,
\begin{equation}
\label{resfin}
\f{d{\hat \sigma}_{a_1a_2}}{d{\hat y} \,dq_T^2} =
\f{d{\hat \sigma}_{a_1a_2}^{(\rm res.)}}{d{\hat y} \,dq_T^2}
+\f{d{\hat \sigma}_{a_1a_2}^{(\rm fin.)}}{d{\hat y} \,dq_T^2} \;\;.
\end{equation}
The logarithmically-enhanced contributions are embodied in the 
`resummed' component $d{\hat \sigma}_{a_1a_2}^{(\rm res.)}$.
The `finite' component $d{\hat \sigma}_{a_1a_2}^{(\rm fin.)}$
is free of such contributions, and it
can be computed by 
truncation 
of the perturbative series at a given fixed order (LO, NLO and so forth).
In practice, after having evaluated $d{\hat \sigma}_{a_1a_2}$ and its 
resummed component at a given perturbative order,
the finite component $d{\hat \sigma}_{a_1a_2}^{(\rm fin.)}$ is obtained by 
the matching procedure described in Sects.~2.1 and 2.4 of 
Ref.~\cite{Bozzi:2005wk}.

The resummation procedure of the logarithmic terms has to be carried out 
\cite{Parisi:1979se}-\cite{Collins:1984kg}
in the impact-parameter space, to correctly take into account the 
kinematics constraint of transverse-momentum conservation.
The resummed component of the partonic cross section 
is then obtained by performing the inverse Fourier (Bessel) transformation 
with respect to the impact parameter $b$.
We write\footnote{In the following equations, the functional dependence on 
the scales $\mu_R$ and $\mu_F$ is understood.}
\beq
\label{resum}
\!\!\! \f{d{\hat \sigma}_{a_1a_2}^{(\rm res.)}}{d{\hat y} \,dq_T^2}({\hat y}, 
q_T,M_H,{\hat s};
\as)
= \f{M^2_H}{\hat s} \;
\int_0^\infty db \; \f{b}{2} \;J_0(b q_T) 
\;{\cal W}_{a_1a_2}({\hat y},b,M_H,{\hat s};\as)
\;,
\eeq
where $J_0(x)$ is the 0th-order Bessel function, and the
factor ${\cal W}$
embodies the all-order dependence on 
the large logarithms $\ln (M_Hb)^2$ at large $b$, which correspond to the
$q_T$-space terms $\ln (M^2_H/q_T^2)$  
(the limit $q_T \ll M_H$ corresponds to $M_Hb \gg 1$, since $b$
is the variable conjugate to $q_T$). 

In the case of the $q_T$ cross section integrated over the rapidity,
the resummation of the large logarithms
is better expressed \cite{Catani:2000vq,Bozzi:2005wk}
by defining the $N$-moments ${\cal W}_N$ of ${\cal W}$
with respect to $z=M^2_H/{\hat s}$ at fixed $M_H$. In the present case, 
where the rapidity is fixed, it is convenient 
(see e.g. Refs.~\cite{Catani:1989ne,Kawamura:2007ze})
to consider `double' $(N_1,N_2)$-moments
with respect to the two variables
$z_1=e^{+{\hat y}} M_H/{\sqrt{\hat s}}$ and 
$z_2=e^{-{\hat y}} M_H/{\sqrt{\hat s}}$ at fixed $M_H$
(note that $0< z_i <1$).
We thus introduce ${\cal W}^{(N_1,N_2)}$ as follows:
\begin{equation}
\label{wnnudef}
{\cal W}_{a_1a_2}^{(N_1,N_2)}(b,M_H;\as) =
\int_0^1 dz_1 \,z_1^{N_1-1} \; \int_0^1 dz_2 \,z_2^{N_2-1} \;\, 
{\cal W}_{a_1a_2}({\hat y},b,M_H,
{\hat s};\as)
\;.
\end{equation}

More generally, any function $h(y;z)$
of the variables $y$ ($|y|< - \ln {\sqrt z}$) and $z$ ($0 < z < 1$) 
can be considered as a function of the two
variables $z_1=e^{+y} {\sqrt z}$ and $z_2=e^{-y}{\sqrt z}$.
Thus, throughout the paper, the $(N_1,N_2)$-moments
$h^{(N_1,N_2)}$ of the function $h(y;z)$ are 
defined as
\begin{equation}
\label{melldef}
h^{(N_1,N_2)} \equiv
\int_0^1 dz_1 \,z_1^{N_1-1} \int_0^1 dz_2 \,z_2^{N_2-1}
\; h(y;z) \;, \quad {\rm where:} \; y=\frac{1}{2} \ln \frac{z_1}{z_2}\;,
\; z=z_1z_2 \;.
\end{equation}
Note that the double Mellin moments can also be obtained 
(see e.g. Ref.~\cite{Sterman:2000pt})
by introducing a Fourier transformation with respect to $y$ 
(with conjugate variable $\nu=i(N_2-N_1)$) and then
performing a Mellin transformation with respect to $z$
(with conjugate variable $N=(N_1+N_2)/2$):
\begin{equation}
\label{n12def}
h^{(N_1,N_2)}=\int_0^1 dz \,z^{N-1} \int_{-\infty}^{+\infty} dy 
\,e^{i \nu y} \;h(y;z) \;, \quad {\rm where:} \; N_1= N+i\nu/2 \;,
N_2= N-i\nu/2 \;. 
\end{equation}

The convolution structure of the QCD factorization formula
(\ref{dcross}) is readily diagonalized by considering $(N_1,N_2)$-moments:
\begin{equation}
\label{n12fact}
d\sigma^{(N_1,N_2)} = \sum_{a_1,a_2} \;f_{a_1/h_1, N_1+1} \;f_{a_2/h_2,N_2+1}
\;d{\hat \sigma}_{a_1a_2}^{(N_1,N_2)} \;,
\end{equation}
where $f_{a/h, N}= \int_0^1 dx \,x^{N-1} f_{a/h}(x)$ are the customary
$N$-moments of the parton distributions.

The use of Mellin moments also simplifies the resummation structure of the
logarithmic terms in  
$d{\hat \sigma}_{a_1a_2}^{(\rm res.) \,(N_1,N_2)}$.
The perturbative factor ${\cal W}_{a_1a_2}^{(N_1,N_2)}$
can indeed be organized in exponential form as follows:
\begin{equation}
\label{wtilde}
{\cal W}^{(N_1,N_2)}(b,M_H;\as) =
{\cal H}^{(N_1,N_2)}(M_H,\as) 
\; \exp\{{\cal G}^{(N_1,N_2)}(\as,{\widetilde L})\}
\;\;,
\end{equation}
where 
\begin{equation}
\label{logdef}
{\widetilde L}= \ln\left(\f{M_H^2b^2}{b_0^2} + 1\right) \;\;,
\end{equation}
$b_0=2e^{-\gamma_E}$ 
($\gamma_E=0.5772\dots$ is the Euler number) and, to simplify the notation,
the dependence on the flavour indeces has been understood.

The structure of Eq.~(\ref{wtilde}) is in close analogy to the cases 
of soft-gluon resummed
calculations for hadronic event shapes in hard-scattering processes
\cite{Catani:1992ua}
and for threshold contributions to hadronic cross sections
\cite{Sterman:1986aj,Catani:1989ne,Catani:1996yz}.
The function ${\cal H}^{(N_1,N_2)}$
(which is process {\em dependent})
does not depend on the impact parameter $b$ and, therefore, its evaluation
does not require resummation of large logarithmic terms. It can be expanded
in powers of $\as$ as
\begin{equation}
\label{hexpan}
{\cal H}^{(N_1,N_2)}(M_H,\as) =
\sigma_{0}(\as,M_H)
\Bigl[ 1+ \f{\as}{\pi} \,{\cal H}^{(N_1,N_2) \,(1)} +
\left(\f{\as}{\pi}\right)^2 
\,{\cal H}^{(N_1,N_2) \,(2)}
+ \dots \Bigr]\;\;, 
\end{equation}
where $\sigma_{0}(\as,M_H)$ is the lowest-order partonic cross section for
Higgs boson production.
The form factor $\exp\{{\cal G}\}$
is process {\em independent}~\footnote{More precisely, it depends only on the
flavour of the colliding partons (see Appendix~\ref{appa}).}; 
it includes the complete dependence on $b$ and,
in particular, it contains 
all the terms that order-by-order in $\as$ are logarithmically
divergent when $b \to \infty$. The functional dependence on $b$ is 
expressed through 
the large logarithmic terms $\as^n {\widetilde L}^m$
with $1\leq m \leq 2n$. 
More importantly,
all the logarithmic contributions to ${\cal G}$ with $n+2 \leq m \leq 2n$
are vanishing.
Thus, the exponent ${\cal G}$ can systematically
be expanded in powers of $\as$, at fixed value of $\lambda=\as {\widetilde L}$,
as follows:
\begin{equation}
\label{gexpan}
{\cal G}^{(N_1,N_2)}(\as,{\widetilde L}) =
{\widetilde L} \,g^{(1)}(\as {\widetilde L})+
g^{(2) \,(N_1,N_2)}(\as {\widetilde L}) +
\f{\as}{\pi} \;g^{(3) \,(N_1,N_2)}(\as {\widetilde L}) + \dots 
\;\;. 
\end{equation}
The term ${\widetilde L} g^{(1)}$ collects the leading logarithmic (LL) 
contributions $\as^n {\widetilde L}^{n+1}$;
the function $g^{(2)}$ resums
the next-to-leading logarithmic (NLL) contributions 
$\as^n {\widetilde L}^{n}$; 
$g^{(3)}$ controls the next-to-next-to-leading logarithmic (NNLL) terms
$\as^n {\widetilde L}^{n-1}$, and so forth.  

Note that we use the logarithmic variable ${\widetilde L}$ 
(see Eq.~(\ref{logdef}))
to parametrize and organize 
the resummation of the large logarithms $\ln (M_Hb)^2$.
We recall the main motivations \cite{Bozzi:2005wk} for this choice.
In the resummation region $M_Hb \gg 1$, we have 
${\widetilde L} \sim \ln (M_Hb)^2$ and the use of the variable 
${\widetilde L}$ is fully legitimate to arbitrary logarithmic accuracy.
When $M_Hb \ll 1$, we have $\tL \to 0$ 
(whereas\footnote{
As shown in Appendix~B of Ref.~\cite{Bozzi:2005wk} (see Eqs.~(131) and (132)
therein), after inverse
Fourier transformation
to $\qt$ space, the $b$-dependent functions $\ln^n (M_Hb)^2$ and
${\widetilde L}^n$ 
lead to 
quite different behaviours at large $\qt$. When $\qt \gg M_H$, 
the behaviour $(1/\qt^2) \ln^{n-1} (\qt/M_H)$ (which is not integrable when
$\qt \to \infty$) produced by 
$\ln^n (M_Hb)^2$ is damped (and made integrable) 
by the extra factor ${\sqrt {\qt/M_H}} \exp (- b_0\qt/M_H)$ produced
in the case of ${\widetilde L}^n$.} 
$\ln (M_Hb)^2 \to \infty$ !) and $\exp \{{\cal G}(\as, \tL)\} \to 1$.
Therefore, the use of ${\widetilde L}$ reduces the
effect produced by the resummed contributions
in the small-$b$ region (i.e., at large and intermediate values of $\qt$), 
where the large-$b$ resummation approach is not justified.
In particular, 
setting $b=0$ (which corresponds to integrate over the entire $\qt$ range)
we have $\exp \{{\cal G}(\as, \tL)\} = 1$: this property
can be interpreted \cite{Bozzi:2005wk}
as a constraint of perturbative unitarity on the total cross section;
the dynamics of the all-order recoil effects, which are resummed 
in the form factor $\exp \{{\cal G}(\as, \tL)\}$,
produces a smearing of the fixed-order $\qt$ distribution of the Higgs boson
without affecting its total production rate.

The resummation formulae (\ref{wtilde}), (\ref{hexpan})
and (\ref{gexpan}) can be worked out
at any given (and arbitrary) logarithmic accuracy
since the functions ${\cal H}$ and ${\cal G}$ can 
explicitly be expressed (see Ref.~\cite{Bozzi:2005wk}) 
in terms of few perturbatively-computable coefficients 
denoted by $A^{(n)}, B^{(n)},
H^{(n)}, C^{(n)}_N, \gamma^{(n)}_N$. The key role of these
coefficients to fully determine the structure of 
transverse-momentum resummation was first formalized by 
Collins, Soper and Sterman \cite{Collins:1984kg, Collins:1981uk,
Catani:2000vq}. 
The present status of the calculation of these coefficients for Higgs boson
production is recalled in Sect.~\ref{sec:lhc}.

In the case of the $q_T$ cross section integrated over the rapidity, 
Eq.~(\ref{wtilde}) is still valid, provided the double 
$(N_1,N_2)$-moments are replaced by the corresponding single $N$-moments
${\cal W}_N, {\cal H}_N, {\cal G}_N$ (see Sect.~2.2 in 
Ref.~\cite{Bozzi:2005wk}). The relation between double and single moments can
easily be understood by inspection of 
Eqs.~(\ref{wnnudef})-(\ref{n12def}).
We see that setting $\nu=0$ in 
Eq.~(\ref{n12def}) 
is exactly 
equivalent to integrate the cross section over the rapidity. Therefore,
the functions ${\cal W}_N, {\cal H}_N, {\cal G}_N$ in 
Ref.~\cite{Bozzi:2005wk} are
obtained by simply setting $N_1=N_2=N$ in the corresponding functions
${\cal W}^{(N_1,N_2)},{\cal H}^{(N_1,N_2)}, {\cal G}^{(N_1,N_2)}$ 
of Eq.~(\ref{wtilde}). 

Moreover, from the results presented in Ref.~\cite{Bozzi:2005wk}, we can 
straightforwardly obtain the functions
${\cal H}^{(N_1,N_2)}$ and ${\cal G}^{(N_1,N_2)}$
from the functions ${\cal H}_N$ and ${\cal G}_N$. 
Roughly speaking, we simply
have
\begin{equation}
\label{n12vsn}
{\cal G}^{(N_1,N_2)} = \frac{1}{2} \left( {\cal G}_{N_1} + 
{\cal G}_{N_2} \right) \;, \quad {\cal H}^{(N_1,N_2)} = \left[ {\cal H}_{N_1}
\; {\cal H}_{N_2}
\right]^{1/2} \;\;.
\end{equation}
More precisely, these equalities are valid in the simplified case where there 
is a single species of partons (e.g. only gluons). In the following we comment
on the physical picture that leads to Eq.~(\ref{n12vsn}). The 
generalization to considering more species of partons
does not require any further conceptual steps: it just involves algebraic
complications related to the treatment of the flavour indeces. 
The multiflavour case is briefly illustrated in Appendix~\ref{appa}.

In the small-$q_T$ (large-$b$) region that we are considering, the kinematics
of the Higgs boson is fully determined by the radiation of soft and collinear
partons from the colliding partons (hadrons) in the initial state.
The radiation of soft partons cannot affect the rapidity of the Higgs bosons.
On the contrary, the radiation of partons that are collinear to $p_1$ ($p_2$),
i.e. in the forward (backward) region, decreases (increases) the rapidity 
of the Higgs boson as a consequence of longitudinal-momentum conservation
(see Eq.~(\ref{kin})).
Since the emissions of collinear partons from $p_1$ and $p_2$ are 
{\em dynamically} uncorrelated (factorized from each other), 
correlations arise only from
kinematics. The use of the $(N_1,N_2)$-moments exactly factorizes 
(see Eqs.~(\ref{dcross}) and (\ref{n12fact})) the kinematical 
constraint of longitudinal-momentum conservation. It follows that the 
$(N_1,N_2)$-dependence of ${\cal W}^{(N_1,N_2)}$ is given by the product
of two functions 
(say,  ${\cal W}^{(N_1,N_2)}= {\cal M}_1^{(N_1)} {\cal M}_2^{(N_2)}$) that
depends only on $N_1$ or $N_2$, respectively.
If all the partons have the same flavour, the two functions should be equal, 
and Eq.~(\ref{n12vsn}) directly follows from 
$[{\cal W}^{(N_1,N_2)}]_{N_1=N_2=N}={\cal W}_N$.

The formalism illustrated in this section defines 
a systematic `order-by-order' (in extended sense)
expansion \cite{Bozzi:2005wk} of Eq.~(\ref{resfin}):
it can be used to obtain predictions with uniform perturbative
accuracy from the small-$q_T$ region to the large-$q_T$ region. The various
orders of this expansion are denoted\footnote{In the literature on $\qt$
resummation, other authors sometime use the same labels (NLL, NLO and so 
forth) with a meaning that is different from ours.}
as LL, NLL+LO, NNLL+NLO, etc., where the 
first label (LL, NLL, NNLL, $\dots$) refers to the logarithmic accuracy at 
small $q_T$ and the second label (LO, NLO, $\dots$) refers to the customary 
perturbative order\footnote{We recall that the LO term at small $\qt$ (i.e. 
including the region where
$\qt=0$) is proportional
to $\as^2$, whereas the LO term at large $\qt$ is proportional to $\as^3$.
This mismatch of one power of $\as$ (and the ensuing mismatch of notation)
persists at each higher order (NLO, NNLO, ...).} at large $q_T$. 
To be precise,
the NLL+LO term of Eq.~(\ref{resfin}) is obtained by including  
the functions $g^{(1)}$, $g^{(2)}$
and the coefficient ${\cal H}^{(1)}$ (see Eqs.~(\ref{gexpan}) and 
(\ref{hexpan}))
in the resummed component, and by expanding the finite (i.e. large-$q_T$)
component up to its LO term. At NNLL+NLO accuracy, the resummed component
includes also the function $g_N^{(3)}$ and the coefficient 
${\cal H}^{(2)}$ (see Eqs.~(\ref{gexpan}) and 
(\ref{hexpan})), while the finite component is expanded up to NLO.
It is worthwhile noticing 
that the NNLL+NLO (NLL+LO) result includes the {\em full} NNLO (NLO)
perturbative
contribution in the small-$\qt$ region.

We recall \cite{Bozzi:2005wk} that, due to
our actual definition of the logarithmic parameter ${\widetilde L}$ in
Eq.~(\ref{wtilde}) and to our matching procedure with the perturbative 
expansion at large $q_T$, the integral over $q_T$ of the $q_T$ cross section
exactly reproduces the customary fixed-order calculation of the total cross
section. This feature 
is not affected by keeping the rapidity
fixed. 
Therefore, the NNLO (NLO) result for total cross section at fixed $y$ 
is exactly recovered upon integration
over $q_T$ of the NNLL+NLO (NLL+LO) $q_T$ spectrum at fixed $y$.

Within our formalism, resummation is directly implemented, at fixed $M_H$,
in the space of the conjugate variables 
$N_1, N_2$ and $b$. 
To obtain the cross
section in Eq.~(\ref{dcross}), as function of the kinematical variables $s,y$ 
and $q_T$, we have to perform inverse integral
transformations. These 
integrals
are carried out numerically.
We recall \cite{Bozzi:2005wk} that the resummed form factor (i.e., each of the
functions $g^{(k)}(\as {\widetilde L})$ in Eq.~(\ref{gexpan})) is singular at 
the value of $b$ where $\as(\mu_R^2) {\widetilde L}=\pi/\beta_0$ ($\beta_0$
is the first-order coefficient of the QCD $\beta$ function). This singularity
has its origin from the presence of the Landau pole in the running of the QCD
coupling $\as(q^2)$ at low scales. When performing the inverse Fourier (Bessel)
transformation with respect to the impact parameter $b$ 
(see Eq.~(\ref{resum})), we deal with this
singularity by using a `minimal prescription' 
\cite{Catani:1996yz,Laenen:2000de}: the singularity is avoided by deforming the 
integration contour in the complex $b$ space (see Ref.~\cite{Laenen:2000de}).
We note that the position of the singularity is completely independent of the
values of $N_1$ and $N_2$. Thus, the inversion of the Mellin moments is 
performed in the customary way (in Mellin space there are no singularities for
sufficiently-large values of Re~$N_1$ and Re~$N_2$). In this respect,
going from single $N$-moments (as in Ref.~\cite{Bozzi:2005wk}) to double
$(N_1,N_2)$-moments (as in the present case, where the rapidity is kept fixed)
is completely straightforward, with no additional (practical or conceptual)
complications.

\section{Higgs boson production at the LHC}
\label{sec:lhc}

In this section we apply the resummation formalism of Sect.~\ref{sec:genfor} 
to the production of the Standard Model Higgs boson at the LHC. 
We closely follow our previous study 
of the single differential (with respect to $q_T$) cross section, with 
the same choice of parameters as stated in Sect.~3 of 
Ref.~\cite{Bozzi:2005wk}.
Therefore, the integration over $y$ of
the double differential 
(with respect to $y$ and $q_T$) cross sections presented in 
this section
returns the $q_T$ cross sections of 
Ref.~\cite{Bozzi:2005wk}.
As a cross-check of the actual implementation of the calculation, 
we have verified that after integration 
over the rapidity the numerical results in Ref.~\cite{Bozzi:2005wk} are 
reobtained within a high accuracy.

As in Refs.~\cite{Catani:2003zt, Bozzi:2005wk},
we use an `improved version' \cite{Kramer:1996iq}
of the large-$M_t$ approximation.
The cross section is first computed by using the large-$M_t$ approximation.
Then, it is rescaled by a Born level factor, such as to include 
the exact lowest-order dependence on the masses, 
$M_t$ and $M_b$, 
of the top and bottom\footnote{We note that the Born level cross section is 
not insensitive to the contribution
of the bottom quark. Adding the bottom-quark loop to the top-quark loop in
the scattering amplitude produces a non-negligible interference effect in
the squared amplitude. The relative effect of the bottom quark decreases
the Born level cross section by about 11\% if $M_H=125$~GeV, and by about
3\% if $M_H=300$~GeV. If $M_H \gtap 500$~GeV, the relative
effect of the bottom quark is
always smaller than 1\%.}
quarks, which circulates in the heavy-quark loop that
couples to the Higgs boson. We use the values $M_t=175$~GeV and $M_b=4.75$~GeV.
As discussed in Ref.~\cite{Catani:2003zt} and recalled in Sect.~\ref{sec:intro},
this version
of the large-$M_t$ approximation is expected to produce an uncertainty that is
smaller than the uncertainties from yet uncalculated perturbative terms
from higher orders.

For the sake of brevity, we present quantitative results only at NNLL+NLO
accuracy, which is the 
highest accuracy that can be achieved by using the present knowledge
of exact perturbative QCD contributions (resummation coefficients and
fixed-order calculations \cite{deFlorian:1999zd}-\cite{Glosser:2002gm}). 
We use the MRST2004 set \cite{Martin:2004ir} of parton distribution functions
at NNLO. The use of NNLO parton densities 
consistently matches
the NNLL (NNLO) 
accuracy of our partonic cross section in the region of small and intermediate
values of $q_T$.

Resummation up to the NLL level is under control from the knowledge of the
perturbative coefficients $A^{(1)}, B^{(1)}, A^{(2)}$ \cite{Catani:vd} and
${\cal H}^{(1)}$ \cite{Kauffman:cx}.
To reach the NNLL+NLO accuracy, the form factor function 
${\cal G}^{(N_1,N_2)}$
in Eq.~(\ref{gexpan}) must
include the contribution from $g^{(3) \,(N_1,N_2)}$ 
(which is controlled by the coefficients $B^{(2)}$ \cite{deFlorian:2000pr}
and $A^{(3)}$ \cite{Vogt:2004mw}),
and the coefficient function ${\cal H}^{(N_1,N_2)}$
in Eq.~(\ref{hexpan}) has to be evaluated up to its second-order term
${\cal H}^{(2) \,(N_1,N_2)}$.
In Ref.~\cite{Bozzi:2005wk} we exploited the unitarity constraint 
${\cal G}(\as,\tL)|_{b=0}=0$ to numerically derive
an approximated form of the coefficient ${\cal H}^{(2)}$ from the NNLO
calculation \cite{NNLOtotal} of the total cross section.
The recent calculation of Ref.~\cite{Catani:2007vq}, which is based
on the complete evaluation of ${\cal H}^{(2) \,(N_1,N_2)}$ in analytic form,
allows us to gauge the
quality of the approximated form. We find that
the use of the ${\cal H}^{(2)}$ of Ref.~\cite{Bozzi:2005wk}
leads to differences of about 1\% with respect to
the exact computation of the rapidity cross section at NNLO.

All the numerical results in this section are obtained by fixing the 
renormalization and factorization scales at the value $\mu_R=\mu_F=M_H$.
The `resummation scale' $Q$ (the auxiliary scale introduced in 
Ref.~\cite{Bozzi:2005wk} to gauge the effect
of yet uncalculated logarithmic terms at higher orders) 
is also fixed at the value $Q=M_H$.
The mass of the Higgs boson is set at the value $M_H=125$~GeV.

\begin{figure}[htb]
\begin{center}
\begin{tabular}{c}
\epsfxsize=10truecm
\epsffile{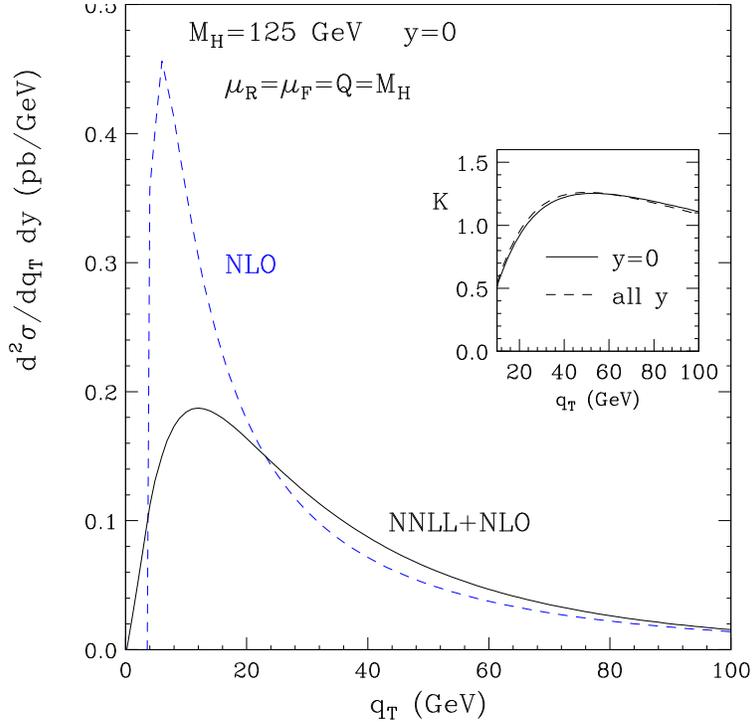}\\
\end{tabular}
\end{center}
\caption{\label{fig:eta0}
{\em 
The $q_T$ spectrum at the LHC with $M_H=125$~GeV and $y=0$: results at NNLL+NLO
(solid line) and NLO (dashed line) accuracy.
The inset plot shows the ratio $K$ (see Eq.~(\ref{kfac}))
of the corresponding $q_T$ cross sections, fixing $y=0$ (solid line) and
integrating them over the full rapidity range (dashed line).} 
}
\end{figure}

We start our presentation of the predictions for Higgs boson production at 
the LHC by considering the $q_T$ dependence of the cross section 
at fixed values of the rapidity. 
In Fig.~\ref{fig:eta0}, we set  $y=0$ and
we compare the customary (when $q_T >0$) NLO calculation (dashed line) 
with the resummed NNLL+NLO calculation (solid line).

As expected, the NLO result diverges to $-\infty$ as $q_T\to 0$ and, at small 
values of $q_T$, it has an unphysical peak that is produced by the numerical 
compensation of negative leading logarithmic and positive 
subleading logarithmic contributions. 
The presence of this peak is not
accidental. At large $q_T$, the perturbative expansion at any fixed order
has no pathological behaviour: it leads to a positive cross section, whose 
value decreases as $q_T$ increases. When $q_T\to 0$, instead, any fixed-order
calculation diverges alternatively to $\pm \infty$ depending on the perturbative
order. Therefore, to go smoothly from the large-$q_T$ behaviour to the 
small-$q_T$ limit, the NLO (or N$^3$LO, and so forth)
calculation of the cross section has to show 
at least one peak in the intermediate-$q_T$ region.

We recall once more that the label NLO in Fig.~\ref{fig:eta0} refers to
(and originates from) the perturbative expansion at large $q_T$. To avoid
possible misunderstandings (coming from such a label) when interpreting the
dashed (NLO) curve in the small-$q_T$ region, we point out that, the only
difference produced in Fig.~\ref{fig:eta0} by the NNLO calculation 
at small $q_T$
(this calculation
can be carried out, for example, by using the NNLO codes 
of Refs.~\cite{Anastasiou:2005qj,Catani:2007vq})
is a spike around the point $q_T=0$. 
More precisely,
{\em as long as} 
$q_T \neq 0$, the dashed curve is {\em exactly} the result of the 
NNLO calculation
of the $q_T$ cross section at small $q_T$. 
The only difference introduced in the plot by
this NNLO calculation 
would
occur in the first bin (with arbitrarily 
small size) that includes the point $q_T=0$. The NNLO value of the 
$q_T$ cross section in this first bin is positive and fixed by the value of the
NNLO total cross section\footnote{By definition,
the integral over $q_T$ of $d^2\sigma/(dq_T \, dy)$
at NNLO is equal to $d\sigma/dy$ at NNLO.}. Of course, owing to the increasingly
negative behaviour of the $q_T$ distribution when $q_T \to 0$, the NNLO value 
of the $q_T$ cross section in the first bin increases by decreasing the
size of that bin.

The resummed NNLL+NLO result in Fig.~\ref{fig:eta0} is physically 
well-behaved at small $q_T$ (it vanishes as $q_T\to 0$ 
and has a kinematical peak at $q_T \sim 12$~GeV), and 
it converges to the expected NLO result 
only when $q_T$ is definitely large ($q_T\simeq M_H$). 

To quantify more clearly the effect of the resummation on the the NLO result,
the value at $y=0$ of the $q_T$ dependent K-factor,
\begin{equation}
\label{kfac}
K(q_T,y)=\f{d\sigma_{NNLL+NLO}/(dq_T \, dy)}{d\sigma_{NLO}/(dq_T \, dy)} \;\;,
\end{equation}
is shown in the inset plot of Fig.~\ref{fig:eta0}. The dashed line shows
the analogous K-factor as computed from the ratio of
the rapidity integrated cross sections.
The similarity between these two K-factors is a first indication 
of the mild
rapidity dependence of the resummation effects. 
By inspection of the inset plot,
we note that NNLL resummation is relevant not only at small $q_T$, but also in
the intermediate-$q_T$ region: as soon as $q_T \ltap 80$~GeV, the resummation
effects are larger than 20\%.
Of course, the fact that $K \sim 1$ at $q_T \sim 24$~GeV is purely accidental:
it simply follows from the unphysical behaviour of the fixed-order perturbative
expansion at small $q_T$.

\begin{figure}[htb]
\begin{center}
\begin{tabular}{c}
\epsfxsize=10truecm
\epsffile{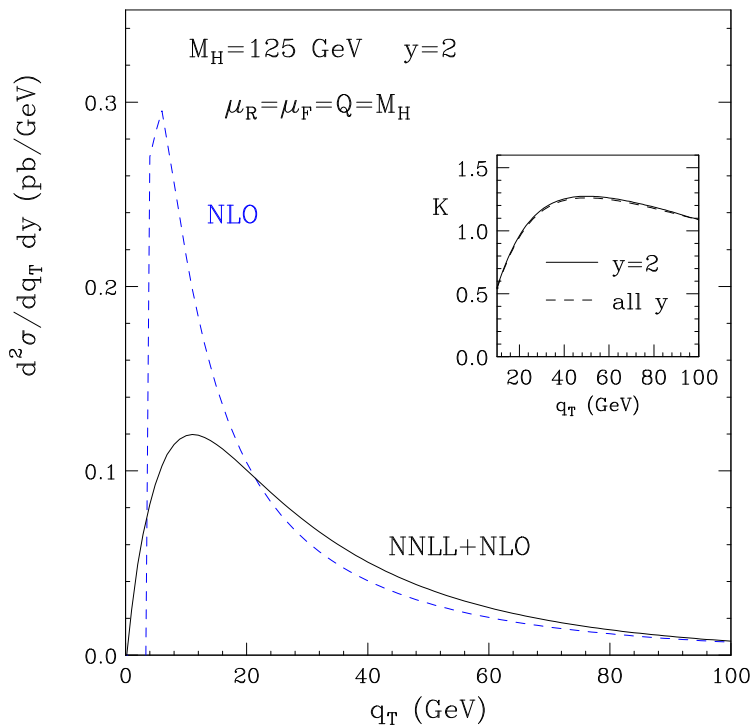}\\
\end{tabular}
\end{center}
\caption{\label{fig:eta2}
{\em The $q_T$ spectrum at the LHC with $M_H=125$~GeV and $y=2$: 
results at NNLL+NLO
(solid line) and NLO (dashed line) accuracy.
The inset plot shows the ratio $K$ (see Eq.~(\ref{kfac}))
of the corresponding $q_T$ cross sections, fixing $y=2$ (solid line) and
integrating them over the full rapidity range (dashed line).}
}
\end{figure}

Considering other values of the rapidity, from the central to the off--central 
rapidity region, we find the same features as observed at $y=0$.
Our results of the $q_T$ spectrum at $y=2$ are presented in 
Fig.~\ref{fig:eta2}. The NNLL+NLO spectrum has a peak at $q_T \sim
11$~GeV.
As happens in the case of the $q_T$ distribution 
integrated
over $y$, the effect of NNLL resummation is definitely non-negligible starting
from relatively-high values of $q_T$. For example, at $q_T=50$~GeV the NNLL+NLO
result is about 30\% higher than the NLO result.

\begin{figure}[htb]
\begin{center}
\begin{tabular}{c}
\epsfxsize=10truecm
\epsffile{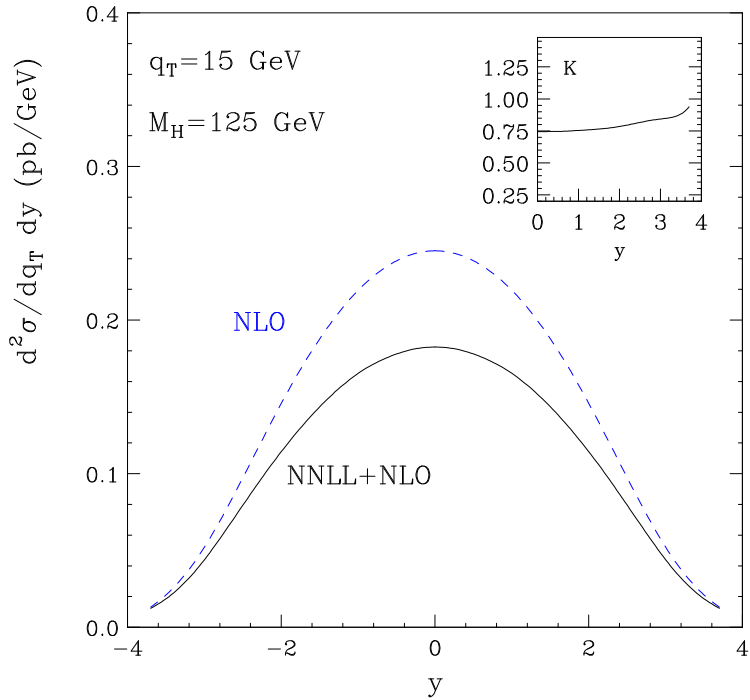}\\
\end{tabular}
\end{center}
\caption{\label{fig:pt15}
{\em 
The rapidity spectrum at the LHC with $M_H=125$~GeV and $q_T=15$~GeV: 
results at NNLL+NLO
(solid line) and NLO (dashed line) accuracy.
The inset plot shows the K-factor as defined in  Eq.~(\ref{kfac}).}}
\end{figure}

\begin{figure}[htb]
\begin{center}
\begin{tabular}{c}
\epsfxsize=10truecm
\epsffile{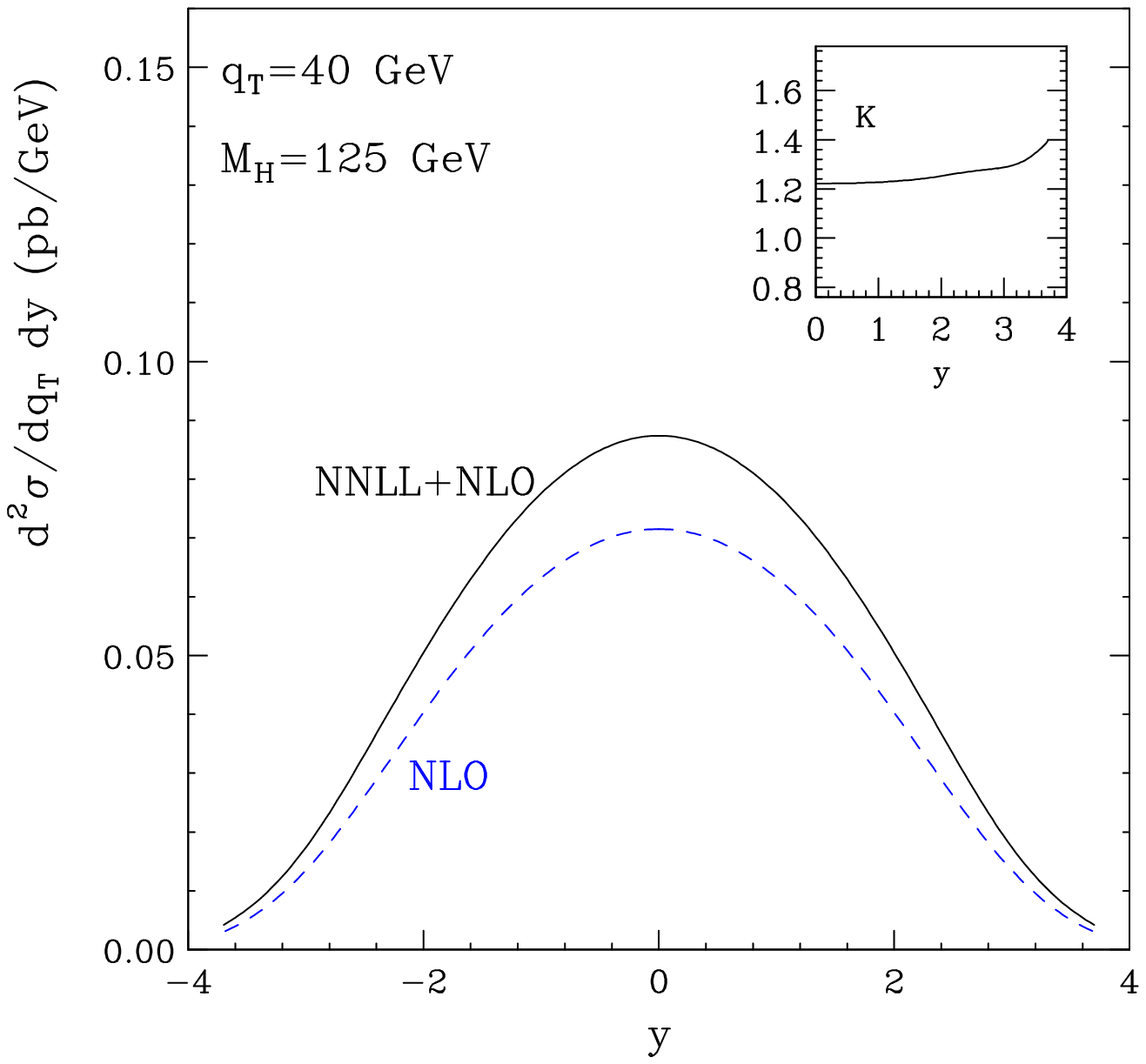}\\
\end{tabular}
\end{center}
\caption{\label{fig:pt40}
{\em 
The rapidity spectrum at the LHC with $M_H=125$~GeV and $q_T=40$~GeV: 
results at NNLL+NLO
(solid line) and NLO (dashed line) accuracy.
The inset plot shows the K-factor as defined in  Eq.~(\ref{kfac}).
}}
\end{figure}

To analyze the rapidity dependence in more detail, we study the 
doubly-differential cross section at fixed values of $q_T$. 
In Figs.~\ref{fig:pt15} and \ref{fig:pt40}, we show quantitative results at two
typical values of the transverse momentum, $q_T=15$~GeV and $q_T=40$~GeV, 
in the small-$q_T$ and intermediate-$q_T$ region, respectively.

Figure~\ref{fig:pt15} shows the rapidity distribution at NNLL+NLO (solid line)
and NLO (dashes) accuracy when $q_T=15$~GeV. At this value of $q_T$,
the effect of NNLL resummation reduces the cross section. 
For example, when $y=0$ the reduction effect is about~25\%.
As can be observed in the inset plot, the relative contribution from the
resummed logarithmic terms is rather
constant in the central rapidity region, and its dependence on $y$ only 
appears in forward (and backward) region, 
where the cross section is quite small.  

When $q_T=40$~GeV (see Fig.~\ref{fig:pt40}), instead, 
the effect of NNLL resummation increases the absolute value of 
the cross section. 
For example, when $y=0$ the NLO cross section is increased by about~22\%.
Nonetheless, as for the relative effect of resummation
and the rapidity dependence of the K-factor, we observe features that are very
similar to those in Fig.~\ref{fig:pt15}.
The resummation effects have
a very mild dependence on $y$ in the central and (moderately) off--central
regions, and this explains the remarkable similarity between the solid and
dashed lines in the inset plot of Figs.~\ref{fig:eta0} and \ref{fig:eta2}.
Since the kinematical region where
$|y| \ltap 2$ accounts for most of the total cross section,
when comparing the ratio $K(q_T,y)$ to the analogous ratio of 
the $y$-integrated cross sections, hardly any differences 
are expected, unless the large-rapidity region is explored.

\begin{figure}[htb]
\begin{center}
\begin{tabular}{c}
\epsfxsize=9truecm
\epsffile{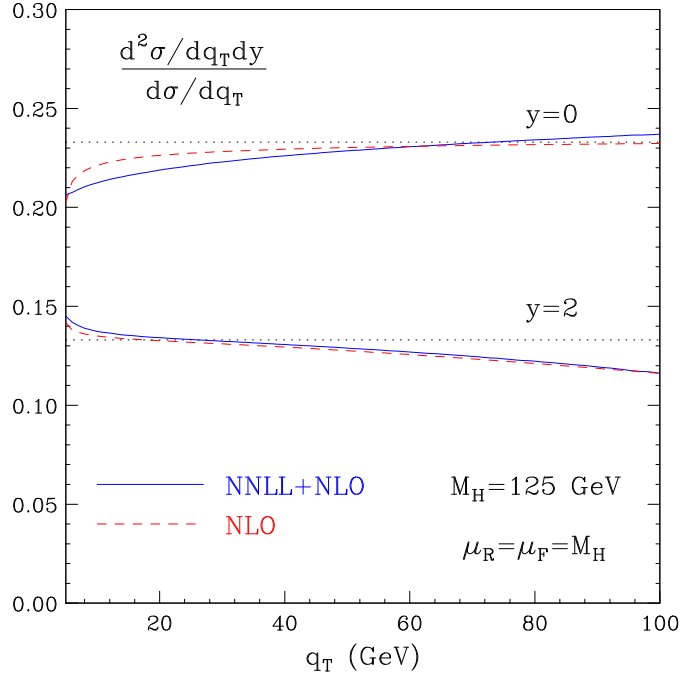}\\
\end{tabular}
\end{center}
\caption{\label{fig:shape}
{\em 
The rescaled $q_T$ spectrum (as defined by the 
ratio 
$R(q_T;y)$
in Eq.~(\ref{rfac})) at the LHC with $M_H=125$~GeV.
The solid (dashed) lines correspond to the NNLL+NLO (NLO) results
at two different values of the rapidity:
$y=0$ (upper) and $y=2$ (lower).
The dotted lines refer to the
corresponding values of the ratio $R_y$ (see Eq.~(\ref{ryfac})). 
  }}
\end{figure}

The mild rapidity dependence of the $q_T$ shape of the resummed results 
can be studied with a finer resolution by defining the following ratio:
\begin{equation}
\label{rfac}
R(q_T;y)
=\f{d^2\sigma/(dq_T \,dy)}{d\sigma/dq_T} \;\;.
\end{equation}
This ratio gives the doubly-differential cross section normalized
to the $q_T$ cross section integrated over the full rapidity range.
For comparison, we consider also the $\qt$-integrated version of the cross
section ratio in Eq.~(\ref{rfac}), and we define the ratio
\begin{equation}
\label{ryfac}
R_y=\f{d\sigma/dy}{\sigma} 
\end{equation}
of the rapidity cross section $d\sigma/dy$
over the total cross
section $\sigma$.

We have computed the ratio in Eq.~(\ref{rfac}) by using the resummed
$\qt$ cross sections at NNLL+NLO accuracy. The results, as a function of $q_T$,
are presented in Fig.~\ref{fig:shape} (solid lines) at two different values, 
$y=0$ and $y=2$, of the rapidity. 
The results of the analogous ($\qt$-independent) ratio $R_y$ 
(computed\footnote{The numerical accuracy of this computation is better than
about 2\%--3\%. Owing to the unitarity constraint in our resummation formalism,
the same result (with a similar numerical accuracy)
can be obtained by integration over $\qt$ of the resummed $\qt$ cross sections.
} at NNLO with the numerical programs
of Refs.~\cite{Anastasiou:2005qj,Catani:2007vq}) at the corresponding 
values of rapidity are also reported (dotted lines) in Fig.~\ref{fig:shape}. 
The dashed lines in Fig.~\ref{fig:shape} correspond to the computation of 
Eq.~(\ref{rfac}) by using the $\qt$ cross sections at NLO: we see that the
dashed and solid lines are very similar (as expected from the similarity
of the dashed and solid lines in the inset plot of Figs.~\ref{fig:eta0} and 
\ref{fig:eta2}).
As discussed below, the results in Fig.~\ref{fig:shape} show that
the cross section decreases and the $\qt$ spectrum softens
when the rapidity increases.

We observe that the lines at $y=0$ lie above the lines at $y=2$;
this is just a consequence of the fact that
the cross sections (both at fixed $\qt$ and after integration over $\qt$)
decrease when $y$ increases.

At fixed $y$, 
$R(q_T;y)$
is not constant: it depends (though very slightly)
on $\qt$.
We note that the corresponding upper and lower 
lines in Fig.~\ref{fig:shape} have different slopes with respect to
$\qt$: fixing $\qt$, the $\qt$ slope of 
$R(q_T;y)$
decreases from positive to negative values as $y$ increases from $y=0$ to
$y=2$, 
thus showing that the $q_T$ spectrum becomes
slightly softer at larger rapidity.

In general, as $|y|$ increases,
the hardness of the $\qt$ shape of $d^2\sigma/(dq_T \,dy)$
decreases.
Since the cross section decreases by increasing the 
rapidity, the hardness 
of $d\sigma/dq_T$ 
(the denominator in Eq.~(\ref{rfac})) is intermediate between the values of the
hardness of $d^2\sigma/(dq_T \,dy)$ (the numerator in Eq.~(\ref{rfac}))
at $y=0$ and at large $|y|$. As a consequence, the $q_T$ slope of 
$R(q_T;y)$
is necessarily positive when $y=0$. 
Note that the $q_T$ slope 
is already negative when $y=2$ (Fig.~\ref{fig:shape}):
this is a consequence of the fact that the bulk of the cross section
is in the rapidity region $|y| \ltap 2$.

Our qualitative 
illustration
of the results in Fig.~\ref{fig:shape} can be
accompanied by some quantitative observations. We note that the rapidity
dependence of the cross sections is sizeable: going from $y=0$ to 
$y=2$, the ratio $R_y$ decreases by 
about 43\%; 
comparable variations affect the
ratio $R(\qt;y)$, which is not very different from $R_y$ and it is slowly
dependent on $q_T$. Indeed, at fixed $y$, the ratio $R(\qt;y)$ at 
NNLL+NLO accuracy has a small and nearly constant slope 
from low values of $\qt$ around the peak
(say, $q_T \sim 10$~GeV) to $q_T=100$~GeV; varying $\qt$
in this region, $R(\qt;y)$ increases by about 11\% when $y=0$, 
and it decreases by about 16\% when $y=2$.
In the same range of $q_T$ and $y$, the values of $R(\qt;y)$ at NNLL+NLO
(solid lines) and at NLO (dashed lines) are very similar: although this is
expected at large $\qt$, the differences
never exceed the level of about 4\% even at values of $\qt$ as low as 
$\qt \sim 10$~GeV.

In summary, the results in Fig.~\ref{fig:shape} show that, 
when $|y|$ increases from the central to the (moderately) off--central region,
the cross sections vary more in absolute value than in $\qt$ shape. 
These features deserve some words of discussion.

We first consider the total cross section $\sigma$ and the rapidity cross
section $d\sigma/dy$. We recall (see Sect.~\ref{sec:intro}) that
the value of these cross sections is sizeably affected by QCD radiative
corrections. The bulk of the effect is due to the radiation of virtual
and soft gluons, and they cannot affect the rapidity of the Higgs boson.
As a consequence, the ratio $R_y$ has little sensitivity to perturbative
QCD corrections. The decreases of $R_y$ as $|y|$ increases is mainly driven 
by the decrease of the gluon density $f_g(x,M_H^2)$ as $x$ increases.
Considering the large-$\qt$ region, similar arguments apply to the 
$\qt$ cross sections $d\sigma/d\qt$ and $d\sigma/(d\qt dy)$, and similar
conclusions apply to the ratio $R(\qt,y)$.
In the small-$\qt$ region, we have to consider the additional and large effect
produced on the $\qt$ cross sections by the logarithmically-enhanced terms
$\ln^m(M_H^2/\qt^2)$. These terms are due to the 
radiation of soft and collinear partons. As already discussed in 
Sect.~\ref{sec:genfor}, the rapidity of the Higgs boson can be varied only by
collinear radiation, while soft radiation can only lead to on overall 
(independent of $y$)  rescaling of the $\qt$ cross sections.
At the LL level, only soft radiation contributes (the LL function $g^{(1)}$
in Eq.~(\ref{gexpan}) does not depend on $N_1$ and $N_2$) and all the
logarithmic terms cancel in the ratio $R(\qt,y)$. The $y$ sensitivity of 
$R(\qt,y)$ starts at the NLL level. 
The corrections produced on the dominant soft-gluon effects by the collinear
radiation are physically \cite{Dokshitzer:hw} well approximated by varying the
scale $\mu$ of the gluon density from $\mu \sim M_H$ to $\mu \sim \qt$.
As a consequence, the variations of the hardness of the of the $\qt$ 
cross sections are mainly driven by
$d \ln f_g(x,\qt^2)/d \ln \qt^2$, the amount of scaling violation of the gluon
density. Since the scaling violation decreases as $x$ increases, the hardness
of $d\sigma/(d\qt dy)$ decreases and the $\qt$ spectrum softens as 
$|y|$ increases. Note that, by increasing $x$, the gluon density decreases 
faster than its scaling violation: this explains why $d\sigma/(d\qt dy)$
varies more in absolute value than in $\qt$ shape when $|y|$ increases.

We conclude this section with some comments about the theoretical uncertainties
on the doubly-differential cross section $d\sigma/(d\qt dy)$ at NNLL+NLO
accuracy. In Ref.~\cite{Bozzi:2005wk} the perturbative QCD uncertainties on 
$d\sigma/d\qt$ were investigated by comparing the results
at NNLL+NLO and NLL+LO accuracies and by performing scale variations at 
NNLL+NLO level. We also considered the inclusion of non-perturbative
contributions, and we found that they lead to small corrections provided 
$\qt$ is not very small.
From these studies we concluded that the NNLL+NLO result has a QCD uncertainty 
of about $\pm 10$\% in the region from small (around the peak of the $\qt$
distribution) to intermediate (say, roughly, $\qt \ltap M_H/3$) 
values of transverse momenta.
Similar studies can be carried out in the case of the doubly-differential cross
section $d\sigma/(d\qt dy)$. These studies are not reported here since their
results and the ensuing conclusions are very similar to those in 
Ref.~\cite{Bozzi:2005wk}. The reason for this similarity is a feature  
that we have pointed out throughout this section: the $\qt$ resummation effects
have a very mild dependence on the rapidity and, thus, they
are almost unchanged when comparing $d\sigma/(d\qt dy)$ with $d\sigma/d\qt$
(equivalently, they largely cancel in the ratio in the cross section ratio 
of Eq.~(\ref{rfac})).

\vspace*{-4.2mm}

\section{Summary}
\label{sec:fin}

We have considered the resummation of the 
logarithmically-enhanced QCD contributions that appear
at small transverse momenta when computing the $\qt$ spectrum of 
a Higgs boson produced in hadron collisions.
In our previous work on the subject
\cite{Bozzi:2003jy,Bozzi:2005wk},
the rapidity of the Higgs boson was integrated over: resummation 
was implemented
by using a formalism based on a 
transform to
impact parameter and Mellin moment space.
In this paper
we have extended the resummation formalism to the case in which the 
rapidity is kept fixed, 
and
we have considered 
the doubly-differential cross section with respect to the 
transverse momentum and the rapidity.
 We have shown that this extension can be carried out without substantial 
complications: it is sufficient to enlarge the conjugate space by introducing
a suitably-defined double Mellin transformation.

The main aspects of our method \cite{Catani:2000vq,Bozzi:2005wk}, 
which are recalled here, are unchanged by the 
inclusion of the rapidity dependence. The resummation is performed at 
the level of the partonic cross section, and the parton densities are 
factorized as in the customary fixed-order calculations.
The formalism is 
completely general and it can be applied to other processes:
the large logarithmic contributions are universal and, thus, they are  
systematically
exponentiated in a process-independent 
form (see Eqs.~({\ref{wtilde}) and (\ref{gexpan}));
the process-dependent part is factorized in the hard-scattering 
coefficient~${\cal H}$.
A~constraint of perturbative unitarity is imposed on the resummed terms
(see Eq.~(\ref{logdef})), so that the $\qt$ smearing produced by the 
resummation does not change the total production rate.
This constraint reduces the effect of 
unjustified higher-order contributions at intermediate 
$\qt$ and facilitates the matching procedure with the complete fixed-order
calculations at large $\qt$. In particular, when the rapidity is kept fixed,
the integration over $q_T$ of $d\sigma/(dq_T dy)$ at NNLL+NLO accuracy 
returns $d\sigma/dy$ at NNLO.

We have presented numerical results for Higgs boson production at the LHC.
Comparing fixed-order and resummed calculations, we 
find 
that the resummation effects are large at small $\qt$ (as expected) and still
sizeable at intermediate $\qt$. The inclusion of the rapidity dependence
has little quantitative impact on this picture since, as we have shown, 
the $\qt$ resummation effects are mildly dependent on the 
rapidity. Going from the central to the (moderately) off--central rapidity 
region, the $\qt$ shape of the spectrum slightly softens.
In the range from small to intermediate values of $\qt$, the residual 
perturbative uncertainty of the 
NNLL+NLO predictions for $d\sigma/(dq_T dy)$
is comparable to that of advanced (NNLO or NNLL+NNLO) calculations
of the $\qt$ inclusive cross sections $d\sigma/dy$ and $\sigma$.

\appendix
\section{Appendix
}
\label{appa}

In this appendix we present the 
structure of the resummation formula
(\ref{wtilde}) by explicitly including the dependence on the flavour 
indeces of the colliding partons. 

In the context of our resummation formalism,
a detailed derivation of
exponentiation in the multiflavour case is illustrated in Appendix~A of
Ref.~\cite{Bozzi:2005wk}.
Considering a generic LO partonic subprocess $c+{\bar c} \to F$
($F=H$ and $c={\bar c}=g$ in the specific case of Higgs boson production
by gluon fusion), and performing $\qt$ resummation after integration over the
rapidity, the resummed component $d{\hat \sigma}_{a_1a_2}^{(\rm res.)}/dq_T^2$
of the partonic cross section is controlled by 
the $N$-moments ${\cal W}_{a_1a_2, \,N}^F$. The final exponentiated result
for these $N$-moments is given by the master formulae (106)--(108)
of Ref.~\cite{Bozzi:2005wk}. We recall the master formula (106) in the following
form:
\begin{equation}
\label{master6}
{\cal W}_{a_1a_2, \,N}^F(b,M;\as)
=\sum_{\{I\}} {\cal H}_{a_1a_2, \,N}^{\{I\}, \,F}\left(M, 
\as\right) \;\exp\{{\cal G}_{\{I\}, \,N}(\as,\tL
)\}
\;\;, 
\end{equation}
where the sum extends over the following set of flavour indices:
\begin{equation}
\label{iset}
\{I\}= c, {\bar c}, i_i, i_2, b_1, b_2 \;.
\end{equation}
and, for simplicity, the functional dependence on various scales (such as
the renormalization and factorization scales) is understood. 
The functions ${\cal G}_{\{I\}, \,N}$ and
${\cal H}_{a_1a_2, \,N}^{\{I\}, \,F}$ are given in the master
formulae (107) and (108), respectively.

In the present paper, $\qt$ resummation is performed at fixed values of the
rapidity, and the double $(N_1,N_2)$-moments
${\cal W}_{a_1a_2}^{(N_1,N_2) \,F}$ in Eq.~(\ref{wnnudef}) replace the 
$N$-moments ${\cal W}_{a_1a_2, \,N}^F$ of Ref.~\cite{Bozzi:2005wk}.
The generalization of Eq.~(\ref{wtilde}) to the multiflavour case
is straightforwardly obtained from Eq.~(\ref{master6}) by the simple 
replacement $N \to (N_1,N_2)$:
\begin{equation}
\label{new6}
{\cal W}_{a_1a_2}^{(N_1,N_2) \,F}(b,M;\as)
=\sum_{\{I\}} {\cal H}_{a_1a_2}^{\{I\}, \, (N_1,N_2) \,F}(M, 
\as) \;\;\exp\{{\cal G}_{\{I\}}^{(N_1,N_2)}(\as,\tL
)\}
\;\;. 
\end{equation}
The exponent ${\cal G}_{\{I\}}^{(N_1,N_2)}$
of the process-independent form factor and
the process-dependent hard factor 
${\cal H}_{a_1a_2}^{\{I\}, \, (N_1,N_2) \,F}$
are
\begin{equation}
\label{calgcap}
{\cal G}_{\{I\}}^{(N_1,N_2)} = {\cal G}_{c} + {\cal G}_{i_1, \,N_1} + 
{\cal G}_{cb_1, \,N_1} + {\cal G}_{i_2, \,N_2} + 
{\cal G}_{{\bar c}b_2, \,N_2} \;\;,
\end{equation}
\begin{equation}
\label{htildeflav}
{\cal H}_{a_1a_2}^{\{I\}, \, (N_1,N_2) \,F}
= \sigma_{c{\bar c}, \,F}^{(0)} \;H_c^{F}
\;S_c \;\,{\widetilde C}_{cb_1, \,N_1} \;
\left[ {\bom E}_{N_1}^{(i_1)} \;{\bom V}_{N_1}^{-1}
\; {\bom U}_{N_1} \right]_{b_1 a_1}
{\widetilde C}_{{\bar c}b_2, \,N_2} \;
\left[ {\bom E}_{N_2}^{(i_2)} \;{\bom V}_{N_2}^{-1} 
\;{\bom U}_{N_2} \right]_{b_2 a_2}
\;\;. 
\end{equation}
The expressions in Eqs.~(\ref{calgcap}) and (\ref{htildeflav})
are completely analogous to the master formulae (107) and (108) 
in Ref.~\cite{Bozzi:2005wk} (the functional dependence on the scales 
$M, \mu_R, \mu_F$ and $Q$ is explicitly denoted in those formulae).
In particular, we note that the dependence of
${\cal G}^{(N_1,N_2)}$ and 
${\cal H}^{(N_1,N_2)}$
on the Mellin variables $N_1$ and
$N_2$ is completely factorized: each of terms on the right-hand side of 
Eqs.~(\ref{calgcap}) and (\ref{htildeflav}) depends only on one
Mellin variable (either $N_1$ or $N_2$). This factorized structure is 
completely consistent with Eq.~(\ref{n12vsn}) and with the physical picture
discussed below Eq.~(\ref{n12vsn});
the dependence on $N_1$ ($N_2$)
follows the longitudinal-momentum flow and the flavour flow 
$a_1\to b_1 \to i_1 \to c$ ($a_2\to b_2 \to i_2 \to {\bar c}$)
that are produced by collinear radiation from the initial-state
parton with momentum $p_1$ ($p_2$).
The various Mellin functions (${\cal G}_{i, \,N}, {\bom E}_{N}^{(i)},
{\bom U}_{N}$ and so forth) in Eqs.~(\ref{calgcap}) and (\ref{htildeflav})
can be found in Ref.~\cite{Bozzi:2005wk}.

\noindent {\bf  Acknowledgements.} 
The work of D.dF. was supported in part
by 
CONICET.
D.dF. wishes to thank the Physics Department of the University of Florence
and INFN
for support and hospitality while this work was completed.

\end{document}